\documentclass{ws-procs9x6}

\begin{document}

\parskip0em

\title{\vspace*{-2.5cm}
A Comparison between an Ultra-Relativistic $Au$+$Au$ Collision and the 
Primordial universe} 

\author{\vspace*{-0.8cm}
Jens S\"oren Lange\footnote{\uppercase{E}mail soeren@bnl.gov.}}

\address{\vspace*{-0.0cm}
Institut f\"ur Kernphysik\thanks{Currently on sabattical leave to \newline
\uppercase{B}rookhaven \uppercase{N}ational \uppercase{L}aboratory, 
\uppercase{U}pton, \uppercase{N}ew \uppercase{Y}ork 11973, \uppercase{USA}}, 
Johann Wolfgang Goethe-Universit\"at,\\
August-Euler-Stra\ss{}e 6, 
60486 Frankfurt/Main, Germany
}

\maketitle

\vspace*{-0.6cm}

\abstracts{
Ultra-relativistic nucleus-nucleus collisions create a
state of matter of high tem\-pe\-ra\-ture and small baryo-chemical potential,
which is similar to the thermodynamical conditions
in the primordial universe. 
Recent analyses of $Au$+$Au$ collisions at RHIC revealed the 
temperature, size and density of the system.
Thus, a comparison to the primordial universe can be attempted. 
In particular, two 
observables shall be investigated, namely
{\it (1)} the temperature at baryon 
freeze-out ($t$$\simeq$10~fm/c in the $Au$+$Au$ collision) and 
{\it (2)} the matter density at the partonic stage ($t$$\leq$1~fm/c). 
}  

\vspace*{-0.6cm}


\linepenalty=10
\widowpenalty=10
\clubpenalty=10

\noindent
At the Relativistic Heavy Ion Collider (RHIC) 
at Brookhaven National Laboratory, New York, USA, 
gold nuclei collide at a maximum center-of-mass energy  
$\sqrt{s}$=200~GeV.
The STAR experiment is one of four experiments [\refcite{rhic_overview}], 
which investigate novel QCD phenomena
at high density and high temperature
in these collisions.
The main STAR subdetector is a midrapidity ($|$$\eta$$|$$\leq$1.6) Time Projection 
Chamber [\refcite{tpc}] (TPC, $R$=2~m, $L$=4~m)  with $\simeq$48,000,000 pixels. 
In 3 years of data taking, high statistics (10$^6$$\leq$$N$$\leq$10$^7$ 
events on tape) for $Au$+$Au$, $p$+$p$ and $d$+$Au$ at $\sqrt{s}$=200 GeV
and $Au$+$Au$ at $\sqrt{s}$=130 GeV were recorded\footnote{The $p$+$p$ and $d$+$Au$ 
data set serve as reference.}. 
Fig.~1 shows a RHIC $Au$+$Au$ collision.

\vspace*{-0.4cm}
\begin{figure}[ht]
\begin{minipage}[b]{45mm}
\centerline{\epsfxsize=45mm\epsfbox{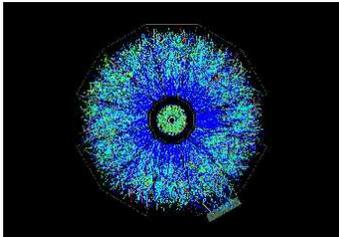}}   
\caption{RHIC $Au$+$Au$ collision at $\sqrt{s}$=200~GeV, recorded at the STAR experiment.}
\label{fevent}
\end{minipage}
\hspace{\fill}
\begin{minipage}[b]{67mm}
\vspace*{-0.7cm}
\section{\mbox{Comparison of the Temperature}}
\label{ctemperature}
\vspace*{-0.1cm}
Generally, in a thermalized system, inclusive particle transverse mass spectra 
can be described by a Boltzmann distribution 
$dN$/$dm_T$$\sim$$m_T$exp(-($m_T$-$m$)/$T$) [\refcite{nic7}],
where $N$ denotes the particle yield.
The transverse mass is defined as $m_T$=$\sqrt{p_T^2+m^2}$.
By using $dN$/$dm_T$ for $\pi^{\pm}$, $K^{\pm}$, protons and anti-protons,
the freeze-out temperature in an $Au$+$Au$ collision was determined 
to be $T$=
\end{minipage}
\end{figure}

\vspace*{-0.55cm}
\noindent
89$\pm$10~MeV [\refcite{identified}], which corresponds
to $T$=(1.03$\pm$0.12)$\cdot$10$^{12}$~K.
 This temperature can be compared
to other systems (Tab.~\ref{ttemperature}). 
An interesting difference is, that an $Au$+$Au$ collision
is matter dominated all the time. In the universe, there has been
at first a radiation dominated time period ($T$$\geq$1~eV), followed 
by a matter dominated period ($T$$\leq$1~eV). Both periods differ in
the time dependence of the temperature, i.e.\ $T$$\sim$1/$t^{1/2}$ vs.\  
$T$$\sim$1/$t^{2/3}$, respectively. For a $Au$+$Au$ collision, the time
dependence can be derived by a simple thermodynamical ansatz $E/N$$\sim$T
and $N$/$V$$\sim$$T^3$$\sim$1/$t$, which leads to a flatter 
time dependence $T$$\sim$1/$t^{1/3}$
(using total energy $E$, particle number $N$ and volume $V$).  
The above mentioned temperature $T_{kin}$=89$\pm$10~MeV refers to
{\it kinetic} freeze-out, i.e.\ the termination of elastic rescattering.
The temperature of the earlier {\it chemical} freeze-out 
(i.e.\ the termination of ineleastic rescattering) is higher 
$T_{chem}$=156$\pm$6~MeV [\refcite{identified}]. 
As it does not depend upon the particle density,  
the {\it chemical} freeze-out temperature is a universal quantity,
i.e.\ identical for the $Au$+$Au$ collisions and the primordial 
universe.

\vspace*{-0.55cm}
\section{Comparison of the Size}
\vspace*{-0.15cm}
\label{csize}

\begin{figure}[hhh]
\centerline{\epsfxsize=3.9in\epsfbox{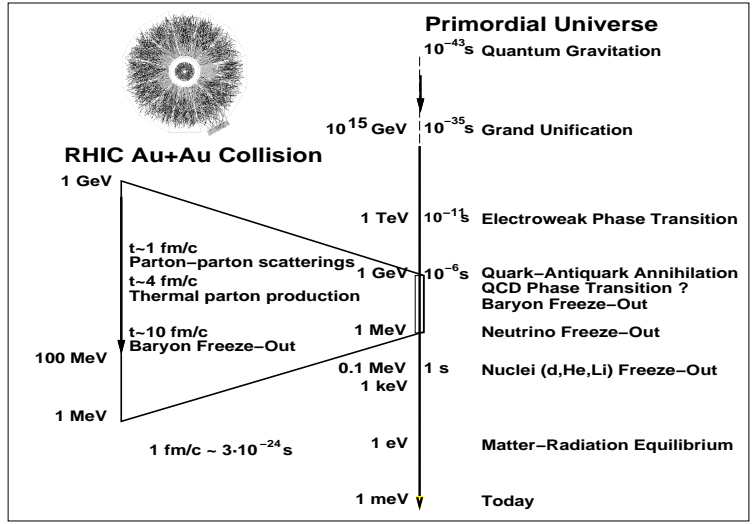}}   
\caption{\hspace*{-0.5em}Timeline of an $Au$+$Au$ collision (left) and the primordial 
universe [5] [6] (right).}
\label{ftime}
\end{figure}

\noindent
The size\footnote{Note that a priori $R_{long}$ and $R_{out}$ are measurements
of homogenity lengths, which then are interpreted as system size.} 
of the fireball in a $Au$+$Au$ collision 
was measured precisely by $\pi^{\pm}\pi^{\pm}$ interferometry [\refcite{hbt}], 
i.e.\ in the beam direction 
$R_{long}$=5.99$\pm$0.19(stat.)\\$\pm$0.36(syst.),
and perpendicular to the beam direction 
$R_{out}$=5.39$\pm$0.18\\(stat.)$\pm$0.28(syst.).
This size corresponds to the time of 
pion freeze-out ($t$$\simeq$10~fm/c).
As the fireball is expanding (Sec.~\ref{cbeta}), the size is a function
of time. 
What was the size of the universe, when it had a temperature of $T$=100~MeV
(corresponding to hadronic freeze-out) ?
As shown in Fig.~\ref{ftime}, hadronic freeze-out is believed to have
occured in the primordial universe at $t$$\sim$10$^{-3}$~s.
As inflation is believed to have occured at
10$^{-36}$~s$\leq$$t$$\leq$10$^{-33}~s$, it can be concluded, that during
baryonic freeze-out the universe had already macroscopic dimensions,
i.e.\ a horizon distance of $R$$\sim$10~km.

\vspace*{-0.55cm}
\section{Comparison of the Expansion Velocity}
\vspace*{-0.15cm}
\label{cbeta}

\noindent
As mentioned in Sec.~\ref{ctemperature}, inclusive particle $m_T$ spectra 
for a $Au$+$Au$ collision can be fitted by a Boltzmann
distribution. This ansatz is only approximately true, as it neglects kinetic terms,
i.e.\ the inverse logarithmic slope is not only proportional to 1/$T$,
but 1/$T$+1/2$m$$v^2$. Only for light particles the approximation $m$$\simeq$0 
holds. If treated more quantitatively [\refcite{identified}], an expansion 
velocity can be extracted from a simultaneous fit to $\pi$, $K$ and $p$
spectra. The result is $\beta$=0.59$\pm$0.05 with $v$=$\beta$c.
For the universe, at the time of baryon freeze-out 
with a horizon distance of $R$$\simeq$10~km
(corresponding to $T$$\simeq$100~MeV), the expansion rate had been 
slower than the speed of light 
by a very large factor 1/$M^2_{Planck}$$\simeq$10$^{19}$ [\refcite{kajantie}].
In addition, the topology of the primordial universe might have been 
very different from the topology of a RHIC collision. 
The general relativity line element $ds^2$$\sim$$dr^2$/(1-$k$) contains 
a curvature $k$=2$G$$M$/$R$$c^2$, using the gravitational constant $G$.
$M$ and $R$ denote the mass and the size of the system, respectively.
Even though the universe appears to be flat ($k$=0) in the present,
the inflation model still allows a primordial non-flat geometry [\refcite{guth}].
For a singularity at $t$=0, 
one would have to assume $k$=+1 for the period before inflation. 
In that case, the universe would have been a 3-dim sphere curved into the 4$^{th}$ 
dimension, and thus infinite for an observer inside the universe.
A particle travelling into one direction,
finally - due to the curvature - comes back to the point, where it started.
The system is closed.
The RHIC collision topology is not a sphere, because 
the two initial state $Au$ nuclei are moving along the beam axis 
with the speed of light. 
Thus, the system has a cylindrical component.
Additionally, in case of the RHIC collision, space time is not curved.
According to [\refcite{disaster}], one can estimate    
$k$=10$^{-22}$. Hence, the system is 
geometrically flat and thermodynamically open,
and any particle can escape.

\vspace*{-0.55cm}
\section{Comparison of the Matter Density}
\vspace*{-0.15cm}

\noindent
For a RHIC $Au$+$Au$ collision, the spatial matter density,
i.e.\ the number of partons per volume 
$\rho$=$N_{partons}$/$V$ 
can be estimated by a Bjorken ansatz [\refcite{Bjorken}].
If assuming that initial and 
final entropy are equal, 
the number of partons at $t$$\leq$1~fm/c  
is equal to the measured number of final state hadrons $N_{hadron}$. 
The volume can be calculated by inserting the fireball radius 
(Sec.~\ref{csize}) into a cylindrical volume (due to Lorentz boost in the beam direction),
i.e.\ $\rho$$\simeq$$dN_{parton}$/$dy$$\cdot$1/($\pi$$R^2$$t$)
using the Lorentz invariant 
rapidity $y$.
For this ansatz, highest densities are
expected early in the collision
(singularity for $t$$\rightarrow$0).
For $t$=0.2~fm/c the matter density is $\rho$$\simeq$20/fm$^3$,
which corresponds to $\simeq$15$\times$$\rho_o$, 
the density of cold gold nuclei.
At this high density, hadrons are definitely non-existent.
In Tab.~\ref{tdensity}, the density is compared to other
systems.
What was the size of the universe at $\rho$=15$\rho_o$ ?
For the scenario of inflation, an initial universe mass $M$$\simeq$25~g is generally
assumed. For inflation from an initial radius of $R_i$=10$^{-40}$~m to a final 
radius of $R_f$=1~m, the density would change from 
$\rho_i$$\simeq$10$^{118}$~kg/cm$^3$ (at $t$$\simeq$10$^{-36}$~s) to
$\rho_f$$\simeq$10$^{-19}$~kg/cm$^3$ (at $t$$\simeq$10$^{-33}$~s).
Thus, RHIC density was achieved {\it during} inflation.
In non-inflation models, the size of the universe in the Planck epoch ($t$$\simeq$10$^{-43}$~s) 
is $R$$\simeq$10$^{-5}$~m. To achieve RHIC density, one would hypothetically\footnote{Even before 
the existence of mass in the context of {\it particle} mass, 
mass can be defined in the context of a {\it gravitational} horizon mass 
$m$=6$c^3$$t$/$G$ [\refcite{Hawking}].}
have to fill it with $m$$\simeq$3600~kg gold. 
If the initial mass were less, RHIC density would never have been achieved.

\vspace*{-0.3cm}
\begin{table}[hhh]
\begin{minipage}[t]{50mm}
\tbl{Comparison of the temperatures in various systems.}
{\footnotesize
\begin{tabular}{@{}ll@{}}
\hline
{} &{}\\[-1.5ex]
1.4$\cdot$10$^{34}$~K & Planck temperature\\[1ex]
1.0$\cdot$10$^{12}$~K & $Au$+$Au$ collision\\[1ex]
10$^9$~K & sun (core)\\[1ex]
15$\cdot$10$^6$~K & supernova\\[1ex]
55$\cdot$10$^6$~K & plasma fusion\\[1ex]
4$\cdot$10$^6$~K & laser fusion\\[1ex]
\hline
{} &{}\\[-1.5ex]
\hline
\end{tabular}
\label{ttemperature} 
}
\end{minipage}
\begin{minipage}[t]{60mm}
\tbl{Comparison of the density in various systems.}
{\footnotesize
\begin{tabular}{@{}ll@{}}
\hline
{} &{}\\[-1.5ex]
2$\cdot$10$^{17}$~kg/cm$^3$ & $Au$ nuclear density \\[1ex]
30$\cdot$10$^{17}$~kg/cm$^3$ & $Au$+$Au$ collision \\[1ex]
$\sim$20,000~kg/cm$^3$ & $Au$ atomic density (solid) \\[1ex]
$\sim$1000~kg/cm$^3$ & metallic hydrogen \\[1ex]
1.1$\cdot$10$^{-26}$~kg/cm$^3$ & universe critical density \\[1ex] 
\hline
{} &{}\\[-1.5ex]
\hline
\end{tabular}
\label{tdensity} 
}
\end{minipage}
\vspace*{-13pt}
\end{table}




\end{document}